\documentclass[11pt]{article}
\usepackage{amsmath}
\usepackage{graphicx}

\def\be{\begin{equation}}
\def\ee{\end{equation}}
\def\bea{\begin{eqnarray}}
\def\eea{\end{eqnarray}}

\def\DRED{\ifmmode{{\rm DRED}} \else{{DRED}} \fi}
\def\DREDD{\ifmmode{{\rm DRED}'} \else{${\rm DRED}'$} \fi}
\def\NSVZ{\ifmmode{{\rm NSVZ}} \else{{NSVZ}} \fi}

\begin{document}

\begin{titlepage}
\begin{flushright}
LTH 717\\
NSF-KITP-06-85\\
hep-ph/0609210\\
\end{flushright}

\vspace*{3mm}

\begin{center}
{\Huge
Anomaly Mediation and Dimensional Transmutation}\\[12mm]

{\bf D.R.T.~Jones}\\

\vspace{5mm} 
Dept. of Mathematical Sciences,
University of Liverpool, Liverpool L69 3BX, UK\\
\vspace{8mm}
{\bf and G.G.~Ross}

\vspace{5mm}
The Rudolf Peierls Centre for 
Theoretical Physics,
Oxford University, 1 Keble Road, Oxford OX1 3NP, UK\\
\end{center}
\vspace{3mm}

\begin{abstract}
We show how a sparticle spectrum characteristic of anomaly mediation 
can arise from a theory whose Lagrangian contains no explicit 
mass scale. The scale of supersymmetry breaking is governed by 
the gravitino mass, which is the vacuum expectation value 
of the $F$-term of the conformal compensator field, and the 
tachyonic slepton problem is resolved by the breaking of a $U'_1$ 
gauge symmetry at a scale determined by dimensional transmutation.

\end{abstract}
\vfill
\end{titlepage}


\section{Introduction}

Anomaly mediation is an attractive alternative to the MSUGRA paradigm for
low energy supersymmetry~\cite{lrrs}-\cite{aprr}. In Ref.~\cite{hjjr} we
showed how augmenting the MSSM with an extra $U_{1}$ broken at high energies
provides a natural solution to the AMSB tachyonic slepton problem if the
extra $U_{1}$ has a Fayet-Iliopoulos term; this model being the explicit
construction associated with the scenario first explored in detail in Ref.~%
\cite{jja}. Here we show that a similar low energy theory can in fact arise
from a \textit{scale invariant} $MSSM\otimes U_{1}^{\prime }$ theory, with
the scale of $U_{1}^{\prime }$ breaking arising by dimensional
transmutation, and the only explicit terms of dimension two and three in the
Lagrangian being those associated with AMSB. The $U_{1}^{\prime }$ breaking
scale also determines the right handed neutrino masses, which in turn
determine the observable neutrino masses via the usual see-saw mechanism.
The low energy theory differs, however, from that of Ref.~\cite{hjjr}, in
that a light MSSM singlet survives the $U_{1}^{\prime }$ breaking; this
raises subtle issues concerning decoupling, as described by Pomarol and
Rattazzi (PR)~\cite{aprr}.

\section{$U(1)^{\prime }$ radiative breaking}

\subsection{The superpotential}

\label{sec:superpot} The superpotential for our theory is 
\begin{equation}
W = W_1 + W_2 +W_3,  \label{eq:spot}
\end{equation}
where $W_1$ contains the Yukawa terms responsible for the quark and lepton
Dirac masses (including the neutrino Dirac masses): 
\begin{equation}
W_1 = Q Y_u t^c H_2 + Q Y_d d^c H_1 + L Y_{e} e^c H_1 + L Y_{\nu} \nu^c H_2
\label{eq:spota}
\end{equation}
and $W_2$ contains additional terms involving an MSSM singlet sector; a pair
of fields $\phi$, ${\overline{\phi}}$ charged under $U^{\prime }_1$, and
gauge singlet fields $U$, $Z$: 
\begin{equation}
W_2 = U(\lambda\phi{\overline{\phi}} - {\textstyle{\frac{{1}}{{2}}}}\rho
Z^2) + {\textstyle{\frac{{1}}{{6}}}}kU^3.  \label{eq:spotb}
\end{equation}
$W_3$ contains terms coupling the two sectors, the purpose of which which
will be to generate the Higgs $\mu$-term and the right-handed neutrino
masses. We will consider two slightly different forms for $W_3$: 
\begin{equation}
W_3^A = \lambda^{\prime \prime }U H_1 H_2 + {\textstyle{\frac{{1}}{{2}}}}%
Y_{\nu^c}\phi \nu^c\nu^c  \label{eq:spotca}
\end{equation}
and 
\begin{equation}
W_3^B = \lambda^{\prime \prime }Z H_1 H_2 + {\textstyle{\frac{{1}}{{2}}}}%
Y_{\nu^c}\phi \nu^c\nu^c.  \label{eq:spotcb}
\end{equation}
Here $\lambda$, $\lambda^{\prime \prime }$, $\rho$, $k$ are coupling
constants while $Y_{u,d,e,\nu}$ and $Y_{\nu^c}$ are $3\times 3$ matrices in
flavour space. The superpotential $W$ is natural in that it contains all
possible cubic terms allowed by the symmetry $MSSM\otimes U^{\prime }_1
\otimes Z_2$, where, under $Z_2$, $Z \to -Z$. and the remaining fields are
invariant (in the case of $W_3^A$) or arranged so that $H_1 H_2 \to - H_1
H_2 $ (in the case of $W_3^B$).

Note that $W$ contains no explicit mass scale. The electroweak scale (and
the associated supersymmetry-breaking scale) are generated via anomaly
mediation, while the $U^{\prime }_1$-breaking scale is generated by
dimensional transmutation, as we shall discuss.

The $U^{\prime }_1$ charge assignments of the various fields are shown in
Table~1.

\begin{table}[tbp]
\begin{center}
\begin{tabular}{|ccccccc|}
\hline
$q_Q$ & $q_{u^c}$ & $q_{d^c} $ & $q_{H_1}$ & $q_{H_2}$ & $q_{\nu^c}$ & $%
q_{\phi} = -q_{{\overline{\phi}}}$ \\ \hline
&  &  &  &  &  &  \\ 
$-\frac{1}{3}q_L$ & $-q_e-\frac{2}{3}q_L$ & $q_e+\frac{4}{3}q_L$ & $-q_e-q_L$
& $q_e+q_L$ & $-2q_L-q_e$ & $4q_L + 2q_e$ \\ 
&  &  &  &  &  &  \\ \hline
\end{tabular}%
\end{center}
\caption{Anomaly free $U_1$ symmetry for arbitrary lepton doublet and
singlet charges $q_L$ and $q_e$ respectively.}
\label{anomfree}
\end{table}

\subsection{The scalar potential}

\label{sec:potmin} Our theory is defined by the superpotential Eq.~(\ref%
{eq:spot}), with the addition of anomaly mediation soft-breaking terms of
the general form 
\begin{eqnarray}
L_{\mathrm{SOFT}} &=&\sum_{\phi }m_{\phi }^{2}\phi ^{\ast }\phi +\left[
m_{3}^{2}H_{1}H_{2}+\sum_{i=1}^{3}\frac{1}{2}M_{i}\lambda _{i}\lambda _{i}+%
\mathrm{h.c.}\right]  \notag \\
&+&\left[ H_{2}Qh_{t}t^{c}+H_{1}Qh_{b}b^{c}+H_{1}Lh_{\tau }\tau ^{c}+\mathrm{%
h.c.}\right],
\end{eqnarray}%
where 
\begin{subequations}
\begin{eqnarray}
M_{i} &=&m_{0}\beta _{g_{i}}/g_{i}  \label{eq:amsba} \\
h_{Y} &=&-m_{0}\beta _{Y}  \label{eq:amsbb} \\
m_{\phi }^{2} &=&{\textstyle{\frac{{1}}{{2}}}}m_{0}^{2}\mu \frac{d}{d\mu }%
\gamma _{\phi }.  \label{eq:amsbc}
\end{eqnarray}%
Here $m_{0}$ is the gravitino mass, $\beta _{g_{i}}$ are the gauge $\beta $%
-functions, $\gamma $ the chiral supermultiplet anomalous dimension, $Y$
stands for any Yukawa coupling and $\beta _{Y_{t,b,\tau }}$ are the Yukawa $%
\beta $-functions. (We will discuss presently the origin of the Higgs $\mu $%
-term and associated soft breaking term).

Let us seek an extremum of the scalar potential such that $\phi, {\overline{%
\phi}}, Z$ obtain vacuum expectation values which are of the same order and
much bigger than $m_0$. We therefore begin by writing down the scalar
potential corresponding to the superpotential $W_2$. Note that this
superpotential has no accidental additional continuous symmetries;
consequently at the extremum there will be no pseudo-Goldstone bosons. The
relevant tree potential may be written 
\end{subequations}
\begin{eqnarray}
V &=& {\textstyle{\frac{{1}}{{2}}}}m_1^2 x^2 + {\textstyle{\frac{{1}}{{2}}}}%
m_2^2 y^2 +{\textstyle{\frac{{1}}{{2}}}}m_3^2 z^2 + {\textstyle{\frac{{1}}{{2%
}}}}m_u^2 u^2 + h_1 uxy + {\textstyle{\frac{{1}}{{2}}}} h_2 u z^2 + {%
\textstyle{\frac{{1}}{{6}}}}h_3 u^3  \notag \\
&+& {\textstyle{\frac{{1}}{{4}}}}\lambda^2 (xy-z^2+ {\textstyle{\frac{{1}}{{2%
}}}}{\overline{k}}u^2)^2 +{\textstyle{\frac{{1}}{{4}}}}\lambda^2 u^2
(x^2+y^2)+{\textstyle{\frac{{1}}{{2}}}}\lambda\rho u^2 z^2 + {\textstyle{%
\frac{{1}}{{8}}}}g_{\phi}^2 (x^2 - y^2)^2.  \label{eq:epot}
\end{eqnarray}
Here $g_{\phi} = q_{\phi}g^{\prime }$ where $g^{\prime }$ is the $U^{\prime
}_1$ coupling, $x /\sqrt{2} = \mathopen\langle \phi\mathclose\rangle $, $y /%
\sqrt{2} = \mathopen\langle {\overline{\phi}}\mathclose\rangle $, $z = \sqrt{%
\frac{\rho}{\lambda}}\mathopen\langle Z\mathclose\rangle $, $u /\sqrt{2} = %
\mathopen\langle U\mathclose\rangle $, $m_{\phi} = m_1$, $m_{{\overline{\phi}%
}} = m_2$, $m_3^2 = 2m_Z^2\lambda/\rho$ and ${\overline{k}}= k/\lambda$. The
cubic couplings $h_{1,2,3}$ are related to $\beta_{\lambda,\rho,k}$
according to Eq.~(\ref{eq:amsbb}): 
\begin{subequations}
\begin{eqnarray}
h_1 & = & {\textstyle{\frac{{1}}{\sqrt{2}}}}h_{\lambda} = -{\textstyle{\frac{%
{m_0}}{\sqrt{2}}}}\beta_{\lambda}  \label{eq:hamsba} \\
h_2 & = & -{\textstyle{\frac{{\sqrt{2}\lambda}}{{\rho}}}}h_{\rho} = {%
\textstyle{\frac{{\sqrt{2}\lambda m_0}}{{\rho}}}}\beta_{\rho}
\label{eq:hamsbb} \\
h_3 & = & {\textstyle{\frac{{1}}{\sqrt{2}}}}h_{k} = - {\textstyle{\frac{{m_0}%
}{\sqrt{2}}}}\beta_{k}.  \label{eq:hamsbc}
\end{eqnarray}

For an extremum we have 
\end{subequations}
\begin{subequations}
\begin{eqnarray}
x\left[ m_{1}^{2}+{\textstyle{\frac{{1}}{{2}}}}g_{\phi }^{2}(x^{2}-y^{2})+{%
\textstyle{\frac{{1}}{{2}}}}\lambda ^{2}u^{2}\right] +y\left[ {\textstyle{%
\frac{{1}}{{2}}}}\lambda ^{2}(xy-z^{2}+{\textstyle{\frac{{{\overline{k}}}}{{2%
}}}}u^{2})+h_{1}u\right] &=&0  \label{eq:minA} \\
y\left[ m_{2}^{2}-{\textstyle{\frac{{1}}{{2}}}}g_{\phi }^{2}(x^{2}-y^{2})+{%
\textstyle{\frac{{1}}{{2}}}}\lambda ^{2}u^{2}\right] +x\left[ {\textstyle{%
\frac{{1}}{{2}}}}\lambda ^{2}(xy-z^{2}+{\textstyle{\frac{{{\overline{k}}}}{{2%
}}}}u^{2})+h_{1}u\right] &=&0  \label{eq:minB} \\
z\left[ m_{3}^{2}-\lambda ^{2}(xy-z^{2}+{\textstyle{\frac{{{\overline{k}}}}{{%
2}}}}u^{2})+\lambda \rho u^{2}+h_{2}u\right] &=&0  \label{eq:minC} \\
u\left[ m_{u}^{2}+{\textstyle{\frac{{1}}{{2}}}}\lambda ^{2}(x^{2}+y^{2})+{%
\textstyle{\frac{{1}}{{2}}}}{\overline{k}}\lambda ^{2}(xy-z^{2}+{\textstyle{%
\frac{{{\overline{k}}}}{{2}}}}u^{2})+\lambda \rho z^{2}\right] &&  \notag \\
+h_{1}xy+{\textstyle{\frac{{1}}{{2}}}}h_{2}z^{2}+{\textstyle{\frac{{1}}{{2}}}%
}h_{3}u^{2} &=&0.  \label{eq:minD}
\end{eqnarray}%
We seek an an extremum such that the $U_{1}^{\prime }$ is broken at a scale
much larger than $m_{0}$. For this to be a minimum, we see from Eq.~(\ref%
{eq:epot}) that it must correspond to $x\sim y\sim z>>u$. Then, setting $%
x\approx y\approx z\approx M$, we have from Eq.~(\ref{eq:minD}) that 
\end{subequations}
\begin{equation}
u\approx -\frac{2h_{1}+h_{2}}{2(\lambda ^{2}+\lambda \rho )},
\label{eq:uvev}
\end{equation}%
so that $u$ is naturally of $O({\textstyle{\frac{{m_{0}}}{{16\pi ^{2}}}}})$.
Now in anomaly mediation we have (for a typical $\hbox{TeV}$ sparticle
spectrum) $m_{0}\sim 40\hbox{TeV}$; we thus obtain, in the case $W_{3}\equiv
W_{3}^{A}$, a contribution to the Higgs $\mu $-term from Eqs.~(\ref%
{eq:spotca}), (\ref{eq:uvev}) of appropriate magnitude (around $0.5\hbox{TeV}
$) without having to assume a small value of the coupling $\lambda ^{\prime
\prime }$. It is also easy to show (from Eqs.~(\ref{eq:minA}), (\ref{eq:minB}%
)) that (to leading order in $M$) 
\begin{equation}
x^{2}-y^{2}=\frac{m_{2}^{2}-m_{1}^{2}}{g_{\phi }^{2}},  \label{eq:xydiff}
\end{equation}%
so that the contribution to the slepton doublet mass is simply 
\begin{equation}
\Delta m_{L}^{2}=\frac{1}{2}g_{\phi }^{2}\frac{q_{L}}{q_{\phi }}%
(x^{2}-y^{2})=\frac{m_{2}^{2}-m_{1}^{2}}{2}\frac{q_{L}}{q_{\phi }}
\end{equation}%
(with a similar formula for the slepton singlet mass) so, as long as we pick
charges such that $q_{L}$ and $q_{e}$ have the same sign (depending on the
sign of $m_{2}^{2}-m_{1}^{2}$), we have a potential solution to the
tachyonic slepton problem. Since $m_{1,2}^{2}$ depend on unknown couplings
we can write 
\begin{equation}
\frac{m_{2}^{2}-m_{1}^{2}}{2q_{\phi }}=\xi
\end{equation}%
where $\xi $ is an effective Fayet-Iliopoulos parameter.

Setting $x = R\cos\Omega$ and $y = R\sin\Omega$, we find also from Eqs.~(\ref%
{eq:minA}), (\ref{eq:minB}) that 
\begin{equation}
\sin 2 \Omega = -\frac{m_3^2 + (2h_1+h_2)u + \lambda\rho u^2}{m_1^2 + m_2^2
+ \lambda^2 u^2}.  \label{eq:omegadef}
\end{equation}
Now our desired extremum corresponds to $\Omega \approx \pi/4$; we see
easily from Eq.~(\ref{eq:omegadef}) and using Eq.~(\ref{eq:uvev}) that at
this extremum we require 
\begin{equation}
\Delta = m_1^2 + m_2^2 + m_3^2 - \frac{(h_1 + {\textstyle{\frac{{1}}{{2}}}}%
h_2)^2}{\lambda(\lambda + \rho)} = 0 + O\left[(m_0^4/M^2)\right].
\label{eq:deltadef}
\end{equation}

Now this looks like a fine-tuning of the fundamental parameters; however
when we bear in mind that all these parameters are functions of scale then
this interpretation changes. The crucial point is that if we are interested
in the effective potential of a set of fields at an (approximately common)
value of these fields much larger than any explicit mass parameters, then
the appropriate scale to evaluate the potential is one equal to the the
scale set by the value of the fields themselves. This simply minimises the
effect of radiative corrections on the tree potential.

Now from Eq.~(\ref{eq:epot}) it is clear that for $m_1^2 + m_2^2 + m_3^2\le
0 $, the potential $V$ is unbounded from below as for $u=0$ and $x = y = z
\to \infty$. Crucially, however, it is quite natural to have $\Delta \le 0$
at some scale (corresponding to $\mu < M$) and then $\Delta > 0$ for $\mu > M
$. We will demonstrate that this mechanism works explicitly in the next
section by examining the variation of $\Delta$ with renormalisation scale.
It follows that the potential will develop a minimum at the scale such that
Eq.~(\ref{eq:deltadef}) is satisfied. It is easy to show from Eq.~(\ref%
{eq:epot}) that in the neighbourhood of the minimum, 
\begin{equation}
V = \frac{1}{2}\Delta M^2 + O\left[m_0^4\right].
\end{equation}
This is, of course, an example of dimensional transmutation as originally
described by Coleman and Weinberg~\cite{coleman}; the relation Eq.~(\ref%
{eq:deltadef}) is analogous to the relation CW wrote down between the
quartic scalar and gauge couplings (valid at the extremum of the potential)
in massless scalar QED. In describing the breaking of a gauge symmetry at
high energies our approach here is reminiscent of Witten's inverted
hierarchy model~\cite{witten}.

The upshot of all this is that without fine-tuning, our scale invariant
potential can quite naturally have an extremum such that $x\sim y\sim z\sim
M>>m_{1,2,3}$, thus achieving $U_{1}^{\prime }$ breaking in a manner similar
to that envisaged in Ref.~\cite{hjjr}, but without the explicit introduction
of the breaking scale in the form of an FI term.

\subsection{The anomalous dimensions}

At one loop the anomalous dimensions of the various fields which only appear
in $W_2$ and $W_3$ are given by (for $W_3 \equiv W_3^A$) 
\begin{subequations}
\begin{eqnarray}
16\pi^2\gamma_{\phi} &=& \lambda^2 + {\textstyle{\frac{{1}}{{2}}}}\mathrm{Tr 
} (Y_{\nu^c})^2 - 2 g^{\prime 2 }q_{\phi}^2  \label{eq:anoma} \\
16\pi^2\gamma_{{\overline{\phi}}} &=& \lambda^2 - 2 g^{\prime 2 }q_{\phi}^2
\label{eq:anomb} \\
16\pi^2\gamma_{U} &=& \lambda^2 + {\textstyle{\frac{{1}}{{2}}}}\rho^2 + {%
\textstyle{\frac{{1}}{{2}}}}k^2 + 2\lambda^{\prime \prime 2}
\label{eq:anomc} \\
16\pi^2\gamma_{Z} &=& \rho^2  \label{eq:anomd}
\end{eqnarray}
while for $W_3 \equiv W_3^B$, Eqs.~(\ref{eq:anomc}) and (\ref{eq:anomd}) are
replaced by 
\end{subequations}
\begin{subequations}
\begin{eqnarray}
16\pi^2\gamma_{U} &=& \lambda^2 + {\textstyle{\frac{{1}}{{2}}}}\rho^2 + {%
\textstyle{\frac{{1}}{{2}}}}k^2  \label{eq:anomce} \\
16\pi^2\gamma_{Z} &=& \rho^2 + 2\lambda^{\prime \prime 2}.  \label{eq:anomf}
\end{eqnarray}
For the MSSM non-singlet fields and $\nu^c$ we have 
\end{subequations}
\begin{eqnarray}
16\pi^2(\gamma_L)^i_j &=& (Y_e)^{ik}(Y_e)_{jk} +
(Y_{\nu})^{ik}(Y_{\nu})_{jk} -2C_H\delta^i_j,  \notag \\
16\pi^2(\gamma_{e^c})^i_j &=& (Y_e)^{ki}(Y_e)_{kj} -2C_{e^c}\delta^i_j, 
\notag \\
16\pi^2(\gamma_{\nu^c})^i_j &=& (Y_{\nu^c})^{ik}(Y_{\nu^c})_{jk}+
2(Y_{\nu})^{ki}(Y_{\nu})_{kj}- 2g^{\prime 2 }q_{\nu^c}^2\delta^i_j  \notag \\
16\pi^2(\gamma_Q)^i_j &=& (Y_d)^{im}(Y_d)_{jm} +
(Y_u)^{im}(Y_u)_{jm}-2C_Q\delta^i_j,  \notag \\
16\pi^2(\gamma_{d^c})^i_j &=& 2(Y_d)^{mi}(Y_d)_{mj} -2C_{d^c}\delta^i_j, 
\notag \\
16\pi^2(\gamma_{u^c})^i_j &=& 2(Y_u)^{mi}(Y_u)_{mj} -2C_{u^c}\delta^i_j, 
\notag \\
16\pi^2\gamma_{H_1} &=& 3(Y_d)^{ij}(Y_d)_{ij} + (Y_e)^{ij}(Y_e)_{ij} +
\lambda^{\prime \prime 2 }-2C_H,  \notag \\
16\pi^2\gamma_{H_2} &=& 3(Y_u)^{ij}(Y_u)_{ij} + (Y_{\nu})^{ij}(Y_{\nu})_{ij}
+\lambda^{\prime \prime 2 }-2C_H,
\end{eqnarray}
where 
\begin{eqnarray}
C_{Q}&=& {\textstyle{\frac{{4}}{{3}}}}g_3^2 + {\textstyle{\frac{{3}}{{4}}}}%
g_2^2 +{\textstyle{\frac{{1}}{{60}}}}g_1^2 + q_{Q}^2 g^{\prime 2},  \notag \\
C_L &=& {\textstyle{\frac{{3}}{{4}}}}g_2^2 +{\textstyle{\frac{{3}}{{20}}}}%
g_1^2 + q_{L}^2 g^{\prime 2}  \notag \\
C_{u^c}&=& {\textstyle{\frac{{4}}{{3}}}}g_3^2 +{\textstyle{\frac{{4}}{{15}}}}%
g_1^2 + q_{u^c}^2 g^{\prime 2},  \notag \\
C_{d^c}&=& {\textstyle{\frac{{4}}{{3}}}}g_3^2 +{\textstyle{\frac{{1}}{{15}}}}%
g_1^2 + q_{d^c}^2 g^{\prime 2},  \notag \\
C_{e^c}&=& {\textstyle{\frac{{3}}{{5}}}}g_1^2 + q_{e}^2 g^{\prime 2},  \notag
\\
C_H &=& {\textstyle{\frac{{3}}{{4}}}}g_2^2 +{\textstyle{\frac{{3}}{{20}}}}%
g_1^2 + q_{H}^2 g^{\prime 2}  \label{eq:Cterms}
\end{eqnarray}
and we have written $q_{H}^2 = q_{H_1}^2 = q_{H_2}^2$, since $q_{H_1} = -
q_{H_2}$.

\subsection{RG evolution and radiative breaking}

In this section we work in the simplified approximation where we neglect $%
\lambda^{\prime \prime }$ and $Y_{\nu}$. As a result the renormalisation
group evolution equations of $\lambda$, $Y_{\nu^c}$, $\rho$ $k$ and $%
g^{\prime }$ become (at one loop) a closed system, which simplifies the
analysis, and makes it the same whether we use $W_3^A$ or $W_3^B$. Moreover
we will assume a simplified form for $Y_{\nu^c}$, to wit 
\begin{equation}
Y_{\nu^c} = \left(\begin{array}{ccc}
0&0&0\\ 0&0&0\\ 0&0&\lambda^{\prime}\\
\end{array}\right).\end{equation}
The $\beta$-functions for the Yukawa couplings are then:

\begin{eqnarray}
16\pi^2\beta_{\lambda} &=& \lambda (\gamma_{\phi}+\gamma_{{\overline{\phi}}%
}+\gamma_U)  \notag \\
&=& \lambda \left[3\lambda^2 + {\textstyle{\frac{{1}}{{2}}}}\lambda^{\prime
2 }+ {\textstyle{\frac{{1}}{{2}}}}\rho^2 +{\textstyle{\frac{{1}}{{2}}}}k^2 -
4g^{\prime 2 }q_{\phi}^2\right]  \notag \\
16\pi^2\beta_{\lambda^{\prime }} &=& \lambda^{\prime
}(\gamma_{\phi}+2\gamma_{\nu^c})  \notag \\
&=& \lambda^{\prime }\left[\lambda^2 + {\textstyle{\frac{{5}}{{2}}}}%
\lambda^{\prime 2 }- 2g^{\prime 2 }(q_{\phi}^2 + 2q_{\nu^c}^2)\right]  \notag
\\
16\pi^2\beta_{\rho} &=& \rho (\gamma_{U}+2\gamma_Z) = \rho\left[\lambda^2 + {%
\textstyle{\frac{{5}}{{2}}}}\rho^2 + {\textstyle{\frac{{1}}{{2}}}}k^2\right]
\notag \\
16\pi^2\beta_k &=& 3k\gamma_U = 3k\left[\lambda^2 +{\textstyle{\frac{{1}}{{2}%
}}}\rho^2 + {\textstyle{\frac{{1}}{{2}}}}k^2\right].  \notag \\
\end{eqnarray}
The $\beta$-function for the $U^{\prime }_1$ gauge coupling is 
\begin{equation}
16\pi^2\beta_{g^{\prime }} = b^{\prime }g^{\prime 3},
\end{equation}
where 
\begin{equation}
b^{\prime }= 3(q_e^2+3(q_{u^c}^2+q_{d^c}^2)+6q_Q^2+2q_L^2+q_{\nu^c}^2)
+2(q_{H_1}^2+q_{H_2}^2)+2q_{\phi}^2.
\end{equation}

We then find using the AMSB mass formula 
\begin{equation}
(m^2)^i{}_j = \frac{1}{2}m_0^2\mu\frac{d}{d\mu}\gamma^i{}_j
\label{eq:amsbmass}
\end{equation}
that 
\begin{eqnarray}
16\pi^2 m_{\phi^2} &=& {\textstyle{\frac{{1}}{{2}}}}m_0^2(
2\lambda\beta_{\lambda} +\lambda^{\prime }\beta_{\lambda^{\prime }}
-4g^{\prime }\beta_{g^{\prime }}q_{\phi}^2)  \notag \\
16\pi^2 m_{{\overline{\phi}}^2} &=& {\textstyle{\frac{{1}}{{2}}}}m_0^2(
2\lambda\beta_{\lambda} -4g^{\prime }\beta_{g^{\prime }}q_{\phi}^2)  \notag
\\
16\pi^2 m_Z^2 &=& {\textstyle{\frac{{1}}{{2}}}}m_0^2( 2\rho\beta_{\rho}).
\end{eqnarray}
Substituting in Eq.~(\ref{eq:deltadef}) we thus have 
\begin{eqnarray}
(16\pi^2)^2\frac{\Delta}{m_0^2} &=& [32 \lambda^5 +16 \lambda^2 k^2 \rho + 8
\lambda \rho^2 k^2 +64 \lambda^4 \rho+4 \lambda^3 \lambda^{\prime 2} +56
\lambda^3 \rho^2  \notag \\
&+&48 \lambda^2 \rho^3 + 8 \lambda^3 k^2 +12 \lambda^2 \lambda^{\prime 2}
\rho +9 \lambda^{\prime 4} \lambda+10 \lambda^{\prime 4} \rho +24 \lambda
\rho^4 +8 \lambda \lambda^{\prime 2} \rho^2  \notag \\
&-&g^{\prime 2} (1024\lambda^2 q_L^2 \rho +256 \lambda^2 q_e^2 \rho -64
\lambda^{\prime 2} q_L^2 \lambda +192 \lambda^{\prime 2} q_L^2\rho -16
\lambda^{\prime 2} q_e^2 \lambda  \notag \\
&+&48 \lambda^{\prime 2} q_e^2 \rho +1024 \lambda q_L^2 \rho^2+256 \lambda
q_e^2 \rho^2 +1024 \lambda^2 q_L q_e \rho -64 \lambda^{\prime 2} q_L q_e
\lambda  \notag \\
&+&192 \lambda^{\prime 2} q_L q_e \rho +1024 \lambda q_L q_e \rho^2) -
g^{\prime 4} (4608q_e^4 \rho+55296 q_L^4 \lambda  \notag \\
&+&38912 q_L^4 \rho +5632 q_e^4 \lambda +97792q_e^2 q_L^2\lambda+73216 q_e^2
q_L^2 \rho +37888 q_e^3 q_L \lambda  \notag \\
&+& 29696 q_e^3 q_L \rho +116736 q_L^3 q_e \lambda +83968 q_L^3 q_e
\rho)]/(8(\lambda+\rho)].
\end{eqnarray}

\includegraphics[angle=-90, width= 5.4in]{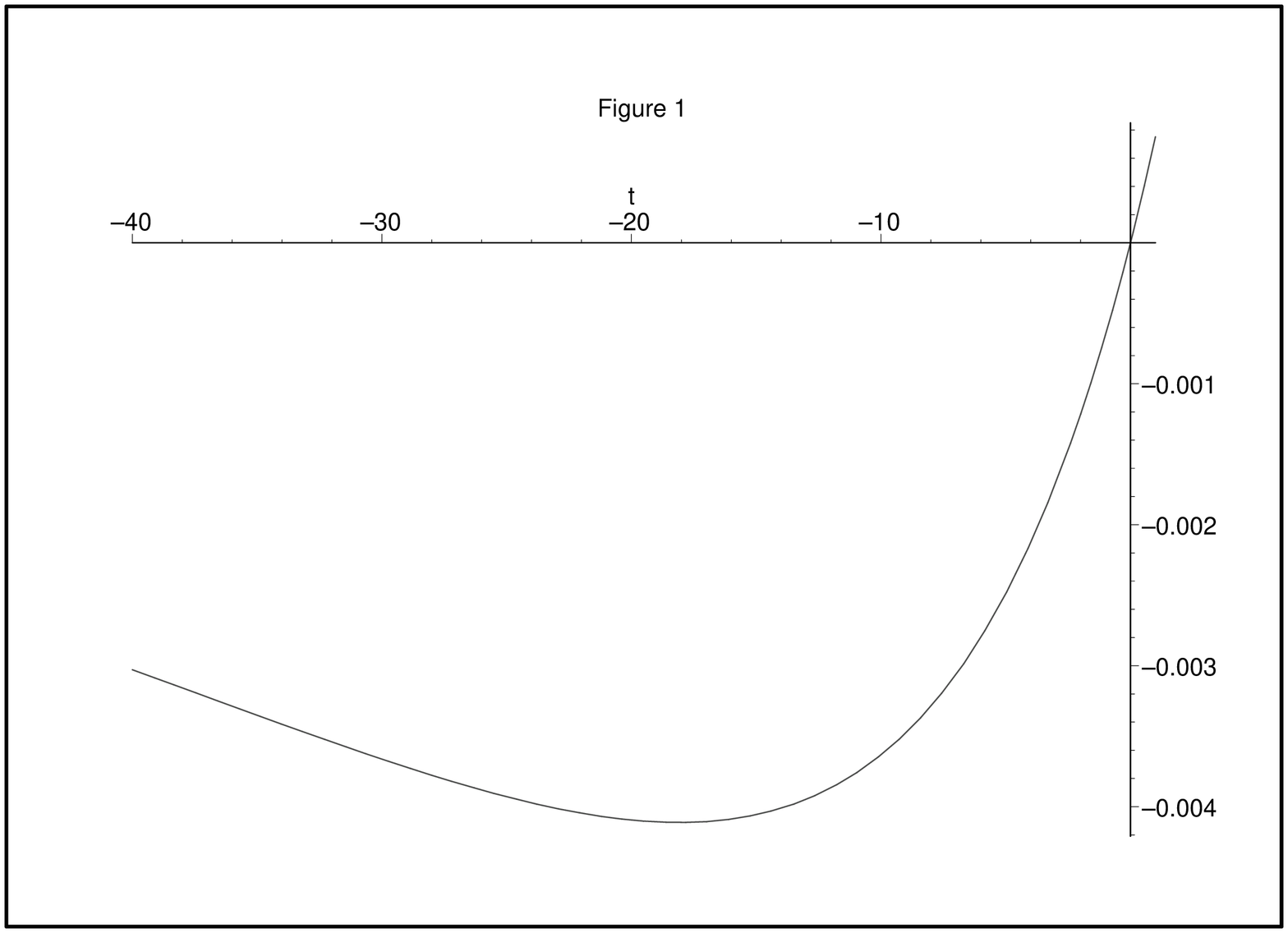} \medskip \leftskip %
= 40 pt\rightskip = 40pt \textit{%
\centerline{ Fig.~1:
Plot of $\Delta$ against $t=\ln {{\textstyle{\frac{{\mu }}{{\mu_{0}}}}}}$}}
\medskip \leftskip = 0 pt\rightskip = 0pt

In Fig.~1 we plot $\Delta $ against $t=\ln {{\textstyle{\frac{{\mu }}{{%
\mu_{0}}}}}}$, choosing initial values so that $\Delta =0$ when $\mu =\mu
_{0}$. (Specifically, we have taken, at $\mu =\mu _{0}$, $\rho =\lambda/10$, 
$\lambda ^{\prime }= k = 2 \lambda $, $\lambda =0.4784$, $g^{\prime }=1$, $%
\xi = 1\hbox{TeV}^2$ and $q_{L}=q_{e}=0.08$. We see that $\Delta$ indeed has
the desired behaviour, with $\Delta <0$ at $\mu <\mu _{0}$ to $\Delta >0$ at 
$\mu >\mu _{0}$. Clearly the potential will have a minimum for $\mu \sim \mu
_{0}$.

\section{The low energy theory}

\subsection{The Higgs spectrum}

If we take the superpotential Eq.~(\ref{eq:spotb}) and shift the fields by
their vacuum expectation values ($U\rightarrow u/\sqrt{2}+U,\phi \rightarrow
x/\sqrt{2}+\phi ,{\overline{\phi }}\rightarrow y/\sqrt{2}+{\overline{\phi }}%
,Z\rightarrow \sqrt{\frac{\lambda }{\rho }}z+Z$), it is easy to see that, to 
$O(\frac{u}{x}),$ the linear combination 
\begin{equation}
H=\frac{1}{R^{\prime }}\left[ x{\overline{\phi }}+y\phi -\sqrt{\frac{2\rho }{%
\lambda }}zZ\right] 
\end{equation}%
where 
\begin{equation}
R^{\prime 2}=x^{2}+y^{2}+{\textstyle{\frac{{2\rho }}{{\lambda }}}}z^{2}
\label{eq:rp}
\end{equation}%
combines with $U$ so that both get large supersymmetric mass terms. The
combination 
\begin{equation}
G=\frac{1}{R}\left[ y{\overline{\phi }}-x\phi \right] 
\end{equation}%
where%
\begin{equation*}
R^{2}=x^{2}+y^{2}
\end{equation*}%
is the Higgs/Goldstone multiplet, combining with the $U_{1}^{\prime }$ gauge
multiplet to produce a spin 1 massive supermultiplet. The orthogonal
combination 
\begin{equation}
L=\frac{1}{R^{\prime \prime }}\left[ x{\overline{\phi }}+y\phi +\sqrt{\frac{%
\lambda }{2\rho }}\frac{x^{2}+y^{2}}{z}Z\right] 
\end{equation}%
where 
\begin{equation}
R^{\prime \prime 2}=x^{2}+y^{2}+\frac{\lambda }{2\rho }\frac{%
(x^{2}+y^{2})^{2}}{z^{2}},  \label{eq:rpp}
\end{equation}%
has a large vacuum expectation value but obtains a mass of order $u$, which
from Eq.~(\ref{eq:uvev}) is of order $\,\hbox{TeV}$. Examined in more
detail, we find that the two scalar components of $L$ have masses of $O(u%
\sqrt{\rho })$ and its fermionic component a mass of order $u\rho $ if, as
we shall motivate later, we choose $\rho <<\lambda $. It is interesting to
note that the $UH_{1}H_{2}$ or $ZH_{1}H_{2}$ terms in the alternative
superpotential terms $W_{3}^{A}$ or $W_{3}^{B}$ of Eqs.~(\ref{eq:spotca})
and (\ref{eq:spotcb}) means that part of our theory resembles the NMSSM.
However, unlike the NMSSM, the light singlet state decouples to $O(\frac{u}{x%
})$ or almost decouples from the MSSM fields, in the first case because $U$
has no $L$ component while in the latter case the coefficient of the
trilinear term must be chosen to be very small for phenomenological reasons.
In both cases however the existence of the light field causes a phenomenon
of AMSB non-decoupling, described by Pomarol and Rattazzi (PR)~\cite{aprr},
which effects the low energy mass spectrum as discussed below.

\subsection{The Higgs $\protect\mu $-term and $b$-term}

\subsubsection{A fine tuned solution}

Let us consider first the case $W_{3}\equiv W_{3}^{A}$ (Eq.~(\ref{eq:spotca}%
)). The vacuum described in Section~\ref{sec:potmin} clearly generates a
Higgs $\mu $-term $\mu _{1}=\lambda ^{\prime \prime }\mathopen\langle U%
\mathclose\rangle $ which is naturally of the right order of magnitude. It
also, however, generates a soft breaking Higgs $b$-term of the form $%
b_{1}=m_{0}\mu _{1}$ which is too large; a generic problem with the scenario
which has been noticed by a number of authors.

There is, however another Weyl invariant operator which generates both $\mu $%
-term and a $b$-term, to wit 
\begin{equation}
\delta L=\sigma \int \,d^{4}\theta \,\frac{\Phi ^{\dagger }}{\Phi }H_{1}H_{2}
\end{equation}%
where $\Phi =1+m_{0}\theta ^{2}$ is the conformal compensator field. This
gives $\mu _{2}=\sigma m_{0}$ and $b_{2}=-m_{0}\mu _{2}$. Thus we have $\mu
=\mu _{1}+\mu _{2}$ and $b=m_{0}(\mu _{1}-\mu _{2})$. It is clear that if we
proceed to determine $\mu $ and $b$ via the electroweak minimisation
conditions in the customary way then this amounts to a degree of fine-tuning
in order to achieve $b\sim \mu ^{2}$

\subsubsection{An unnatural solution}

PR~\cite{aprr} propose an ingenious solution to this problem as follows.
They introduce the further contribution to the superpotential 
\begin{equation}
W_{4}=\lambda ^{\prime \prime }SH_{1}H_{2}+{{\frac{{1}}{{6}}}}k^{\prime
}S^{3}+{{\frac{{1}}{{2}}}}k^{\prime \prime }ZS^{2}.  \label{eq:spotd}
\end{equation}
The superpotential $W_4$ produces at one loop kinetic mixing of $Z$ and $S$: 
\begin{equation}
\delta L \sim k^{\prime }k^{\prime \prime} \int\, d^4 \theta\, Z^{\dagger} S.
\end{equation}
The large vev for $Z$ produces a large supersymmetric mass term for $S$, so
that it decouples at low energies and can be eliminated via its equation of
motion which for small $k^{\prime }$ is 
\begin{equation}
S = - k^{\prime \prime }\frac{H_1 H_2}{Z}
\end{equation}
giving rise to a contribution to the effective low energy theory of the form 
\begin{equation}
\delta L \sim \int\, d^4 \theta\, \frac{Z^{\dagger}}{Z}H_1 H_2
\end{equation}
and hence via anomaly mediation a contribution to the $\mu$ term $\mu =
\mu_3 $ and an associated contribution to the $b$-term of the form 
\begin{equation}
b_3 = -m_0 \beta_{\mu_3}.  \label{eq:prmu}
\end{equation}
Thus one achieves a $b$ of the right order of magnitude (relative to $\mu_3$%
). Then if $\lambda^{\prime \prime }$ and $\rho$ were sufficently small that
the contributions to $b$ associated with them were of similar size to $b_3$
then we would have $\mu \sim \mu_3$ and $b \sim b_1 + b_2 + b_3$, so that
once more we can determine $\mu, b$ via electroweak minimisation. (If we
assume that the PR mechanism is the only source of $\mu, b$ then it is
difficult to achieve the electroweak vacuum~\cite{jja}, because Eq.~(\ref%
{eq:prmu}) would mean $\mu$ and $b$ were not independent.)

The problem with this scenario is that in order to write down the
superpotential $W_{4}$ we must abandon the $Z\rightarrow -Z$ symmetry which
rendered our theory natural. This is because the simultaneous presence of
the $S^{3}$ term and the $ZS^{2}$ term requires $Z$ and $S$ have the same
symmetry properties. As a result a natural theory would also allow
additional terms in Eq.~(\ref{eq:spotb}) of the form $Z^{3}$ and $Z^{2}S$
which unfortunately spoil the mechanism proposed by PR.

\subsubsection{A natural solution without fine tuning}

Here we propose a very simple mechanism which does not involve additional
fields to generate the $\mu $ and $b$ terms of the correct magnitude in a
way consistent with the $Z_{2}$ symmetry. This is achieved through the case $%
W_{3}\equiv W_{3}^{B}$ (Eq.~(\ref{eq:spotcb})).

The $U^{\prime }_1$-breaking proceeds exactly as before and so from Eq.~(\ref%
{eq:spotcb}) we obtain a $\mu$-term of the form $\mu = \lambda^{\prime
\prime }\mathopen\langle Z\mathclose\rangle $. Since $\mathopen\langle Z%
\mathclose\rangle $ is of order the $U^{\prime }_1$ breaking scale it is
therefore necessary to assume that $\lambda^{\prime \prime }$ is very small.
We will return to this point presently.

The reason that we can achieve a suitable $b$-term is that since 
\begin{equation}
Z=\frac{1}{\sqrt{R^{\prime 2}+R^{\prime \prime 2}}}\left( -R^{\prime
}H+R^{\prime \prime }L\right),
\end{equation}%
we can, c.f. Eqs.~(\ref{eq:rp}) and (\ref{eq:rpp}), arrange by making $\rho
\ll \lambda $ that $Z\sim L$. This will suppress the $B$ term since while $%
F_{H}= \mathopen\langle H\mathclose\rangle m_{0}$, $F_{L}=0$. Quantitatively
we have 
\begin{equation}
b=\sqrt{2}\frac{\rho }{\lambda }\mu m_{0}.
\end{equation}%
Thus we retain naturalness by avoiding the need for cancellation between
distinct contributions to $b$. Although we do require two dimensionless
couplings ($\rho $ and $\lambda ^{\prime \prime }$) to be small this can be
achieved by, for example, the Froggatt-Nielsen mechanism, generating these
terms through higher dimension terms suppressed by a large messenger mass.

\subsection{The Low Energy Theory}

The existence of a light singlet field, $L,$ introduces additional
contributions to the soft supersymmetry breaking terms due to the
non-decoupling phenomenon discussed in PR \cite{aprr}. Following PR we have,
in place of Eqs.~(\ref{eq:amsba}), (\ref{eq:amsbb}) and (\ref{eq:amsbc}),
the equations 
\begin{subequations}
\begin{eqnarray}
M_{i} &=&m_{0}\beta _{g_{i}}/g_{i} \\
h_{Y} &=&-m_{0}\left( \frac{\partial }{\partial \ln \mu }+\frac{\partial }{%
\partial \ln M}\right) \ln Z_{Y}(\mu ,M) \\
m_{\phi }^{2} &=&{-{\frac{{1}}{{2}}}}m_{0}^{2}\left( \frac{\partial }{%
\partial \ln \mu }+\frac{\partial }{\partial \ln M}\right) ^{2}\ln Z_{Y}(\mu
,M),
\end{eqnarray}
where $M$ is the $U(1)^{\prime }$ breaking scale and $\mu $ is the
low-energy scale parameter ($\mu \ll M).$ The $\mu $ dependent terms are the
normal anomaly mediated terms while the $M$ dependent terms are the
non-decoupling terms due to the light scalar. These equations apply in the
limit $\rho \ll \lambda $ where $F_{L}=0.$ One sees that the effect of these
non-decoupling terms is simply to retain the contributions of the $%
U_{1}^{\prime }$ gauge boson to the soft breaking terms even though the $%
U_{1}^{\prime }$ gauge coupling freezes out at the breaking scale $M.$ Thus
for example we have for the soft $Qt^{c}H_{2}$ coupling, $h_{t}$, 
\end{subequations}
\begin{equation}
16\pi ^{2}h_{t}=-m_{0}\lambda _{t}\left( 6\lambda
_{t}^{2}-2(C_{H}+C_{Q}+C_{t^{c}})\right),
\end{equation}%
where we have for simplicity retained only the top-quark Yukawa coupling.
The gauge pieces $C_{Q}$ etc. are as given in Eq.~(\ref{eq:Cterms}), but the 
$U_{1}^{\prime }$ coupling is to be evaluated at the $U_{1}^{\prime }$
breaking scale.

\subsection{Phenomenological implications}

For both the cases $W_{3}\equiv W_{3}^{A}$ and $W_{3}\equiv W_{3}^{B},$ in
the limit $g^{\prime }\rightarrow 0,$ the additional soft terms due to
non-decoupling associated with the light scalar state are negligible and, to
a good approximation, we recover the spectrum of Ref.~\cite{hjjr} for the
MSSM states. This is because, as discussed earlier, the coupling of the
light singlet $L$ to the Higgs is very small. The fermionic component, $%
\widetilde{L},$ of $L$ will, however be the LSP. The accelerator lower
limits of around $46\,\hbox{GeV}$ will not apply because of the weak
coupling of the state; in our favoured scenario for Higgs $\mu $-term
generation we estimate its mass to be around $10\,\hbox{GeV}$. The
cosmological and phenomenological implications of a $\widetilde{L}$ LSP are
similar to that of a gravitino LSP and will be discussed in detail
elsewhere. 


For the case when the new gauge interaction is not negligible there will be
additional supersymmetry breaking soft contributions to the MSSM states. We
will also return elsewhere to a consideration of the phenomenology in this
case too.

\section*{\protect\large Acknowledgements}

Both DRTJ and GGR visited CERN and KITP (Santa Barbara) 
while part of this work was done. This
work was partially supported by the EC 6th Framework Programme
MRTN-CT-2004-503369, and by the National Science Foundation  
under Grant No. PHY99-07949.


\begin{thebibliography}{9}
\bibitem{lrrs} L. Randall and R. Sundrum, \emph{Nucl. Phys.} B 557 (1999) 79

\bibitem{glmr} G.F. Giudice et al, \textit{JHEP\/} 9812 (1998) 27

\bibitem{aprr} A. Pomarol and R. Rattazzi, \textit{JHEP \/} 9905 (1999) 013

\bibitem{hjjr} R.~Hodgson, I.~Jack, D.R.T.~Jones and G.G. Ross, \emph{Nucl.
Phys.} B 728 (2005) 192

\bibitem{jja} I.~Jack and D.R.T.~Jones, \emph{Phys. Lett.} B 482 (2000) 167 

\bibitem{coleman} S.R. Coleman and E. Weinberg, \emph{Phys. Rev.} D 7 (1973)
1888

\bibitem{witten} E. Witten, \emph{Phys. Lett.} B 105 (1981)267 
%
%
%
\end{thebibliography}
\end{document}